\documentclass[lettersize,journal]{IEEEtran}

\usepackage{amssymb}
\usepackage{graphicx}
\usepackage{amsmath}
\usepackage{array}
\usepackage{float}
\usepackage{caption}
\usepackage{subcaption}
\usepackage{xcolor}
\usepackage{multirow}
\usepackage{algorithm}
\usepackage{algorithmic}
\begin{document}
\title{An Ensemble Scheme for Proactive Dominant Data Migration of
Pervasive Tasks at the Edge}

\author
{
Georgios Boulougaris, Kostas Kolomvatsos
\thanks{G. Boulougaris and K. Kolomvatsos are with the Department of Informatics and Telecommunications, University of Thessaly, 35100, Lamia, Greece {\tt\small \{gboulougar, kostasks\}@uth.gr}
}
}

\maketitle

\begin{abstract}
Nowadays, a significant focus within the research community on the intelligent management of data at the confluence of the Internet of Things (IoT) and Edge Computing (EC) is observed. In this manuscript, we propose a scheme to be implemented by autonomous edge nodes concerning their identifications of the appropriate data to be migrated to particular locations within the infrastructure, thereby facilitating the effective processing of requests. Our objective is to equip nodes with the capability to comprehend the access patterns relating to offloaded data-driven tasks and to predict which data ought to be returned to the original nodes associated with those tasks. It is evident that these tasks depend on the processing of data that is absent from the original hosting nodes, thereby underscoring the essential data assets that necessitate access. To infer these data intervals, we utilize an ensemble approach that integrates a statistically oriented model and a machine learning framework. As a result, we are able to identify the dominant data assets in addition to detecting the density of the requests. A detailed analysis of the suggested method is provided by presenting the related formulations, which is also assessed and compared with models found in the relevant literature. 
\end{abstract}


\IEEEpeerreviewmaketitle

\section{Introduction}
Nowadays, we are witnessing the provision of a huge infrastructure 
as the combination of two individual ecosystems, i.e.,
the Internet of Things (IoT) and Edge Computing (EC).
This combined infrastructure assists in advancing 
legacy Pervasive Computing (PC) that targets to spread computing devices around end users and 
support them with constantly available computational resources \cite{Kurkovsky}.
The ultimate goal is to adopt the network-connected devices and support new opportunities for pervasive data management and processing. 
IoT devices could be responsible to directly interact with users or their environment and collect the ambient data while EC nodes
could act as mediators between the IoT ecosystem and the Cloud back end being capable of performing various processing activities to support
applications.
Both types of devices are close to data sources and users, shaping the new form of \textit{Pervasive Edge Computing} (PEC) \cite{najam}. Pervasive applications can now be developed more easily since IoT devices interact seamlessly with the environment and users, while EC provides the necessary data processing tools. 
For sure, the collected data can be transferred to the Cloud for additional processing, typically as part of long-term decision-making, as all the discussed devices/nodes in both ecosystems are characterized by limited computational capabilities. 
Apparently, this imposes various challenges mainly related to the type of processing provided (e.g., real-time) and the amount of data stored in the EC nodes. 
No matter than massive resources could be deployed at the network edge to evolve it into a large distributed computing environment, it would still not be an appropriate place for processing large volumes of data \cite{najam}.

Multiple challenges should be met in PEC environments especially related with the management of data and the available services. 
Both, are adopted to support the execution of the desired processing activities requested in the form of tasks.
The need is to have PEC providing on-node applications that focus on real-time data processing, and therefore, each node should be prepared to host distributed datasets using the local resources. 
This preparation involves deciding which services and data should be stored locally and which should be transferred to the Cloud (decision that affects the storage of the collected data) or other nodes. 
It becomes clear that this decision is influenced by the tasks received by a node, which come with specific constraints related to the required services and data. 
If the necessary services and data are not available, nodes can either offload the task or ask for the migration of the required services and data at the local repository \cite{kolomvatsostnsm}, \cite{kolomvatsosFGCS}.
Tasks offloading and data/services migration are solutions to the same problem met when a node is not capable of executing a requested task because either the required data/services are absent or the current load is high.
In any case, the offloading of a task should be supported by a mechanism that selects the appropriate peer to host it or decides to transfer the task to the Cloud.
The former scenario focuses on the attempt to keep the tasks as much as possible in the EC ecosystem, thus, reduce the latency in the provision of responses, while the latter scenario is adopted subject to the rejection of the former one.

If we focus on the offloading of tasks to peer nodes at the EC ecosystem, it becomes clear that 
the new hosting node (the one that receives an offloaded task) should own the appropriate data for executing the offloaded tasks and exhibit a load that leaves the room for this additional processing activity.
The new hosting node could identify if requests for data processing are repeated and, in order to alleviate its load, it could 
`push' those data to the requesting nodes. 
Apparently, this approach demands an intelligent scheme for deciding \textit{which data assets} should be transferred to the requesting nodes.
The goal is to support a \textit{proactive data migration model} based on the `push' mode with the new hosting node being the actor that initiates the process.
In this paper, we propose a scheme for detecting the data upon which the offloaded tasks request for processing and initiate 
an interaction with the requestor to migrate those data and minimize the offloaded activities. We assume that nodes adopt a `myopic' behaviour, i.e., when they do not own the appropriate data or the load is high, they offload the task to the correct peer without any other further processing.
Actually, any node in the ecosystem could adopt our model which can also be extended
to arm EC nodes with the ability of `planning' data migration at their neighbors based on the received offloaded tasks.
It concerns a very dynamic scenario as data and tasks are continually updated due to the needs of applications.

\textbf{\textit{Motivating Scenario}}. We can focus on a scenario where a sink node is responsible to collect data from a neighborhood (we consider that those data fit to the local resources). We assume that the sink node owns better computational resources than the other members of the group but less resources than the Cloud. In any case, the sink node can host a significant part of the aggregated data reported by group members.  
Apparently, group members own a set of local distributed datasets, however, those datasets are continually updated as new data arrive. Hence, nodes should evict some data towards the sink node and the Cloud.
In this scenario, the sink node becomes the receptor of offloaded tasks to respond with the outcomes of a processing upon the aggregated data.
This is natural, as nodes cannot host the entire set of aggregated data to avoid 
a high number of replicas and save resources. 
However, all nodes receive requests for processing upon a data interval being the result of the interaction of IoT devices with end users. 
When data are not locally present, the task is offloaded to the sink node which owns the aggregated dataset.
It becomes obvious that the sink node, to save resources and reduce the load, could detect the data requested by nodes and initiate a `push' migration action.

Our model is built upon an ensemble scheme which involves two technologies, i.e.,
the Multivariate Kernel Density Estimation (MKDE) scheme \cite{hardle} and an One Class Support Vector Machine (OCSVM) model \cite{SVM_for_ND}.
Both techniques are applied upon the data requests of the offloaded tasks.
The MKDE model can expose information about the structure in data while the OCSVM scheme can identify instances of a particular class based on a training dataset containing only instances of that class.
MKDE is a non-parametric approach to estimate the probability density function of random variables and OCSVM
works in an unsupervised manner to classify every input data based on the adopted objective 
function and generates a label as an output to specify if that data are outliers or not.
A comparison of their performance for the same problem is provided by \cite{ngan}.
We strategically rely on the ensemble scheme as the OCSVM model
cannot calculate the density of data focusing on the model of the decision boundary between samples considered as `natural' values or outliers.
The performance of the OCSVM is combined with the detected density of the data requests provided by the MKDE scheme.
The result is that we  can efficiently detect the data being the `core' requests of the offloaded tasks considering them as parts of a potential migration activity.
The salient contributions of the paper are as follows:
\begin{itemize}
  \item A mechanism to detect the density of the data requested by the offloaded tasks based on the MKDE model; 
  \item A scheme for detecting the outliers and the strong data subspaces adopting the OCSVM model;
  \item A decision making mechanism that combines the aforementioned models to deliver the final data subspaces that become parts of a migration activity; 
  \item A comprehensive evaluation of our mechanisms against baselines showcasing the benefits and its applicability.
\end{itemize}

In the following lines, we present the organization of the paper.
We report on the related work in Section \ref{related} and present the 
basic information around our problem in Section \ref{overview}.
The description of the proposed approach is performed 
in Section \ref{mechanism}.
In Section \ref{setup}, we discuss the outcomes of the 
adopted experimental evaluation and conclude this paper in  
Section \ref{conclusions} by giving some of the envisioned 
future research plans.

\section{Prior Work}
\label{related}
The combination of the IoT and EC infrastructures clearly results in a layered architecture that may allow for the allocation of data and services at any node present there. Starting form the bottom layer, IoT devices can collect data while interacting with the environment and end users, then, those data are reported through streams to the Cloud back end. The intermediate layer, i.e., the EC ecosystem, can relay the collected data and keep some of them locally to support various processing activities. In 
\cite{kolomvatsosfusion}
one can find example efforts that focus on the management of the collected data towards the provision of localized services and decision making models. 
The processing of the incoming streams is a critical research subject when we want to 
deliver real-time decisions while the processing of data at rest is employed for long-term decisions. We rely on the concept of tasks, i.e., any processing activity requested to be executed over a set of data. Example tasks could be 
queries requesting data retrieval \cite{elzeki}, \cite{Hsieh}, the training of Machine Learning (ML) models 
\cite{kolomvatsosFGCS}, data selection for caching \cite{ref32}, video pre-processing \cite{ref34} and so on and so forth.
Tasks are executed by a node that is characterized by its load, computational capabilities, communication interfaces, the local dataset, etc.
Apparently, the efficient management of the incoming tasks/data and the speed of processing will affect the throughput of every node. 
When nodes are overloaded or do not own the requested data, they can transfer/offload tasks to peers or the Cloud, as described in \cite{kolomvatsosFGCS}, \cite{wang}. Nodes can also rely on services or data migration, as mentioned in \cite{kolomvatsostnsm}, \cite{kolomvatsosfgcs2}, 
\cite{li2}, to address gaps in their capabilities and avoid malfunctions or delays in the provision of responses. An intelligent approach is to combine all these actions and enhance the capabilities of EC and IoT nodes keeping data and processing as close to end-users as possible. 
Additionally, to achieve a high level of collaboration between autonomous nodes, one can adopt various strategies, such as the aggregation of locally possessed data \cite{ref17}, the collaboration between nodes \cite{ref21}, or having a sink node that stores data and performs processing activities as the single representative of a group 
\cite{ref26}. Regardless of the chosen strategy, the problem remains complex when nodes should act independently to meet the performance challenges posed by dynamic and distributed environments.

By analyzing the behavior of nodes, we can detect the need for the efficient management of incoming tasks and available data. Both tasks offloading and data migration/replication deal with the same problem, i.e., the effective service of processing requests. 
As nodes have the choice to offload tasks and data to the Cloud, the preferred option is to keep them within the EC ecosystem, thus, minimizing the latency in the provision of the final outcomes. 
However, task offloading decisions face multiple challenges such as the management of the communication overhead \cite{Breitbach}, the knowledge on the data present at peers 
\cite{kolomvatsostkde}, 
the load of peers \cite{kolomvatsosFGCS} and so on and so forth. 
To achieve the best performance, any decision-making mechanism should incorporate the relevant contextual information as noted before.
A map of the available data in the ecosystem can be maintained at every node to guide the offloading activities when necessary \cite{kolomvatsosfgcs2}, while a monitoring mechanism of the nodes' status can trigger offloading actions \cite{boulougaris}. 
Concerning the technologies adopted for offloading activities, the research community has proposed various mechanisms including cooperative schemes \cite{ref2}, swarm intelligence technologies \cite{razavinegad}, \cite{yang}, genetic algorithms \cite{hu}, graph-based schemes and their intelligent management \cite{coltin}, and optimal stopping theory \cite{Ouyang}. 
Any proposed model should consider the combination of the local load and the demand for each task before delivering the final action 
\cite{kolomvatsosFGCS}. 
Nodes could benefit from being aware of the distribution, trends, and access patterns of incoming tasks over their local data. 

To avoid the overloading of the network, service migration can be performed in parts by automating the process of connecting service pieces to the hosting infrastructure. ML and particularly reinforcement learning, can improve the decision-making for service migration \cite{bellavista2}. 
However, the adoption of supervised ML models can be challenging as we have to rely on a representative dataset that incorporates the appropriate distribution of the present data \cite{alam}, \cite{wang2}.
For instance, it will be difficult to collect data that incorporate all the possible 
versions of nodes' status as one can observe an increased level of uncertainty
in the corresponding decision making.
Additionally, service migration models can also be formulated as a queue stability control problem \cite{Ouyang1} that can be solved through optimization schemes like Lyapunov or multi-objective optimization \cite{sun}, \cite{xu1}. 
Past research efforts mainly focus on minimizing the latency while the consideration of energy consumption constraints in EC setups should also play a crucial role. 
In any case, user-centric service migration models take more time to optimize unless some assumptions are made or approximate solutions are foreseen. 
Markov chains can be used to solve the problem, but they could require a long time to converge and reach their stationary distribution \cite{mosel}. 
In \cite{wang1}, 
a finite-state Markov decision process
through the utilization of a modified policy-iteration algorithm. 
In \cite{Ouyang2}, the authors present a Thompson-sampling based model for dynamic service migration decisions while in \cite{sun1}, another service migration scheme is proposed for networked vehicles supported by services 
located at the Cloud. 

The efficient utilization of computing resources present at the EC infrastructure could be also achieved through data migration and replication enabling the local processing 
while filling the gaps in the datasets
and eliminates data transfer overheads. 
To avoid latency, data should be placed at the appropriate nodes in a proactive manner
\cite{kolomvatsostkde}. 
Replication is useful when multiple data consumers are located in different places 
and exhibit interest on the same data \cite{huang}. 
Data can be transferred through \textit{push} or \textit{pull} modes, or by a central entity that places data in the right locations. 
Edge to edge learning can optimize the organization of the edge ecosystem in terms of 
organization \cite{kumar}. 
Optimization models can eliminate delays in data transmission and computation time
\cite{le}, 
making them useful for applications that transfer data within the network. 
Methods for data collection that support privacy and secure data channels to resist injection attacks are also proposed \cite{zhang}. 
The communication overhead plays a significant role in data transfer, as observed in high-security models for transferring large-scale data to the Cloud and inter-cloud knowledge migration schemes \cite{rao}. 

In this paper, we propose a model for data migration, however, focusing on a `push' model.
We enhance the autonomous behaviour of nodes that have to administrate incoming tasks as the result of 
offloading decision from other peers.
We assume that these offloading decisions are made due to the absence of the appropriate data to respond to processing requests.
We also assume that peers offloading tasks are not armed with a intelligent mechanism 
to detect the required data.
Hence, every node receiving the offloaded tasks can identify the data they desire to execute processing activities and 
select if it is going to push them to the requestors.
Apparently, the decision of pushing data is followed by a short interaction between nodes in order to 
conclude the activity.
The node making the `push decision' can select a sub-set of data or neighboring peers (close to the node offloading tasks) to migrate the decided information when the peer sending the offloading tasks cannot host the entire set of data.
This approach significantly deviates from the respective literature as 
it is not referring to a centralized framework that decides that placement of data in the ecosystem and 
focuses on the needs of nodes selected to execute the offloaded tasks.

\section{Problem Description \& Preliminaries}
\label{overview}
In Table \ref{nomenclature}, we provide the notations adopted throughout this paper. 
We focus on the PEC ecosystem where numerous nodes can be present and elaborate on a collaborative approach 
for the management of processing tasks and the collected data. Collaborative activities may have the form of knowledge, services or data exchange as well as collaborative processing and tasks offloading 
\cite{hu}, \cite{kolomvatsostnsm}, \cite{kolomvatsosFGCS}.
Managing tasks is a significant research challenge that affects the performance of nodes and the applications and services that end users rely on. Tasks demand for local data processing that may involve statistical reasoning, inferential analytics, and real-time data management, 
e.g., estimation of top-$k$ lists over the incoming 
data streams.
As users and applications interact with IoT and EC ecosystems, they require data processing, thus, indirectly indicating the tasks and services that need to be executed or invoked. 
Tasks offloading involves the selection of a peer node to allocate a task for execution for various reasons, such as the lack of relevant data, the increased local workload, or the absence of appropriate services. 
If we focus on the data aspect of the problem, we can easily identify that the creation of a data map based on the received requests plays a vital role in enhancing nodes' productivity. 
By analyzing the map, we can identify the data that should be present in each local dataset, 
and the remaining data can be transferred upwards to the Cloud backend to support other long-term processing activities.
It becomes obvious that the analysis of the incoming tasks can lead to the aforementioned map and a model that places the collected data at the appropriate nodes.
For instance, we can select the data kept at the appropriate PEC nodes and avoid as much as we can offloading actions 
that may increase the load of the new receptors, fire new offloading actions and so on and so forth.
As the `appropriate' nodes, we define nodes that are instructed to execute specific tasks upon specific data assets.
In this paper, we propose a model for assisting edge nodes to detect the data that peer nodes demand when they offload their tasks 
and fire a push scheme that interacts with the requestors to migrate the local data to them.
If the processing requests (i.e., tasks) for a specific data interval 
are continuous, this should trigger the receptor to initiate the data migration process to the requestor.

\textbf{Definition}. Requestor is a node that offloads a task to a peer in the EC ecosystem requesting for processing.

\textbf{Definition}. Receptor is a node that receives an offloaded task from its peers. 

It should be noticed that, receptors are selected to host the offloaded tasks as they may are the hosts of the demanded data like it is presented in \cite{kolomvatsosFGCS}.
On the other side, requestors may decide to offload a task as may not host the appropriate data and 
may not identify the need for migrating them locally due to the absence of an intelligent scheme for administrating the local dataset or may not want to host such data due to resource constraints. 
Moreover, this difficulty in executing a task may be temporal in the sense that 
resources or relevant data can be available in the near future (e.g., reception of new data assets, release of the currently taken resources and so on and so forth).

We target to a set of nodes $\mathcal{N} = \left\lbrace n_{1}, n_{2}, \ldots, n_{N}\right\rbrace$ having the capabilities to interact with peers and IoT devices while maintaining local datasets
that consist of multivariate 
vectors $\mathbf{x} = [x_{1}, x_{2}, \ldots, x_{M}]^{\top} \in \mathbb{R}^{M}$
($M$ is the number of dimensions). 
Apart from data, nodes also receive requests for tasks execution in the form of
$\mathcal{T}=\left\lbrace \left[ y^{l}_{j}, y^{h}_{j} \right] \right\rbrace$,
$j = \left\lbrace 1, 2, \ldots, M \right\rbrace$ with the indexes $l$ and $h$ representing
the lowest and the highest value for the $j$th dimension, respectively. 
Tasks define the range of data upon which the demanded processing activities should be executed.
Apparently, $y^{l}_{j} \leq y^{h}_{j}$.
For simplicity and without loss of generality, in our analysis, we focus on a specific 
dimension, thus, the notation becomes $\mathcal{T}=\left\lbrace \left[ y^{l}, y^{h} \right] \right\rbrace$,
$y^{l}, y^{h} \in \mathbb{R}^{+}$ (negative values can be also adopted but omitted in this analysis).
The stream of tasks define a list of data requests with the maximum range being 
annotated by 
$\left[ \min{y^{l}}, \max{y^{h}} \right]$.
We have to notice that different peer nodes, through the offloading of their tasks, impose different 
data intervals for the processing activities. Hence, every node should maintain 
a set of `threads' to monitor the requests coming from peers and detect the data missing from their local datasets.

\begin{figure}[h]
\centering
\includegraphics[width=0.35\textwidth]{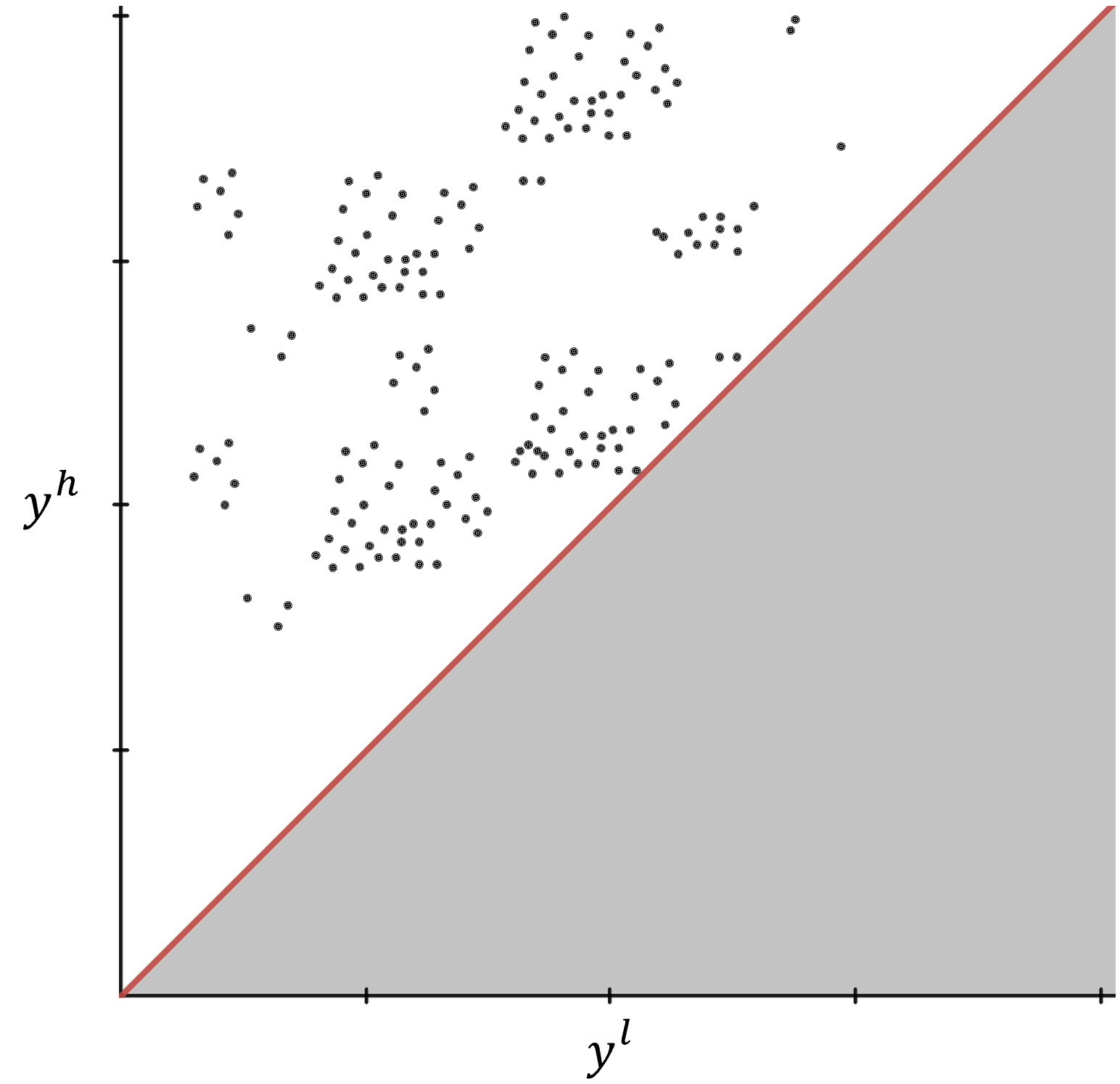}
\caption{Plot for data requests as demanded by potential tasks.}
\label{fig1}
\end{figure}

Apparently, the reception of an offloaded task that demands the processing upon 
$\left[ y^{l}, y^{h} \right]$ triggers a matching activity with the
available data $\mathbf{x}$. If the requested data are locally present, the processing could start after placing the task to the local queue.
If the data are absent or the load of the receptor is high, the incoming tasks could be the subject of an additional offloading process.
For instance, $\left[ 23, 33 \right]$ ($y^{l}_{j} = 23$, $y^{h}_{j} = 33$) could be part of a select query 
asking for data being in the specific interval and some processing upon them.
The interesting is that due to the presence of nodes in a very dynamic environment
where data and requests for tasks are continuously updated, the statistics
of $\mathbf{x}$ and $\left[ y^{l}, y^{h} \right]$ are subject to change.
This observation depicts the difficulty in maintaining the entire set of 
the collected data due to the limited storage and computational resources.
In Figure \ref{fig1}, we plot an example of $\left[ y^{l}, y^{h} \right]$ pairs of the requested data
for a specific dimension. Similar plots can be exposed for the additional dimensions and peers offloading tasks.
We cannot meet any data assets in the grey area as $y^{l} \leq y^{h}$.

Without loss of generality, we focus on a specific peer (requestor) and the interaction with the receptor node which receives the offloaded tasks.
We get a stream of requests depicting the offloaded tasks in a time window $W$
$\left[ y^{l}_{t}, y^{h}_{t} \right]$, $t=1,2, \ldots, W$. 
Actually, we have a time series for this pair of values $\left[ y^{l}_{t}, y^{h}_{t} \right]$ upon which we have to identify 
if and what data should be pushed to the requestor.
Our intention is to detect the `frequent' requested data, i.e., the data requests that dominate the time series and, then, propose their migration to the requestor.
We strategically select to be based on an ensemble scheme that combines
the known MKDE \cite{hardle}
and an ML algorithm, i.e.,
an OCSVM model \cite{SVM_for_ND}.
The outcome is the `dominant' 2D data vectors that have the highest probability to 
be requested in the future.
The estimation of those data pairs is, then, compared with the local data and 
those become subject to a migration action.



\begin{table}[h!]
\centering
\caption{Nomenclature}
\begin{tabular}{ll}
\hline\hline
Notation & Description \\
\hline\hline
   $N$ & Number of nodes \\
   $\mathcal{N}$  &  Set of nodes  \\
   $n_{i}$  &  The $i$th node \\
   $\mathbf{x}$ & Multivariate data vector \\
   $M$  &  The number of dimensions in multivariate vectors \\
   $\mathcal{T}$ & The request for processing \\
   $\left[ y^{l}_{j}, y^{h}_{j} \right]$ & Requested interval for processing \\ 
   $W$ & The time window where we process the offloading requests \\
\hline\hline       
\end{tabular}
\label{nomenclature}
\end{table}

\section{Tasks Aware Data Detection and Selection}
\label{mechanism}
In this section, we present an analytical description of the proposed ensemble scheme for deciding the 
dominant data pairs in the incoming processing requests as exposed by the offloaded tasks.
For each part of the scheme, we provide the appropriate formulations, then, we discuss the final 
migration model built on the push mode as already explained. 

\subsection{Statistically Oriented Data Detection}
\label{datamap}
In this Section, we described the adopted MKDE upon two variables, i.e., $y^{l}, y^{h}$ for every dimension of the available data.
In general, Kernel Density Estimation (KDE) can be extended to estimate multivariate densities $f$ in $\mathbb{R}^{p}$ (in our case $p=2$) based on the same principle \cite{García-Portugués}: perform an average of densities `centered' at the data points. 
The generic description is as follows. For a sample $X_{1}, X_{2}, \ldots, X_{n}$ in $\mathbb{R}^{p}$, the KDE of $f$ evaluated at $x \in \mathbb{R}^{p}$ is defined as
\begin{equation}
\label{mKDE}
\hat{f}(x;H)=\frac{1}{n|H|^{1/2}}\sum_{i=1}^{n} K(H^{-1/2}(x-X_{i})),
\end{equation}
where $K$ is the \textit{multivariate kernel}, a $p$-variate density that is (typically) symmetric and unimodal at zero, and that depends on the bandwidth matrix $H$, a $p \times p$ symmetric and positive definite matrix.
A common notation for the Kernel function is $K_{H}(z)=|H|^{-1/2}K(H^{-1/2}z)$, which is the so-called scaled kernel, so the KDE can be compactly written as 
\begin{equation}
    \hat{f}(x;H)=\frac{1}{n}\sum_{i=1}^{n} K_{H}(x-X_{i})
\end{equation} 
The most employed multivariate kernel is the Normal Kernel $K(z)=\phi(z)=(2\pi)^{-p/2}e^{-\frac{1}{2}z'z}$, for which $K_{H}(x-X_{i})=\phi_{H}(z-X_{i}$. 
Then, the bandwidth $H$ can be thought of as the variance-covariance matrix of a multivariate normal density with mean $X_{i}$ and the KDE can be regarded as a data-driven mixture of those densities.
The meaning of Eq(\ref{mKDE}) is similar to that of the Univariate KDE, i.e., to construct a mixture of densities, where each density is centered at a respective data point. Consequently, and in broad terms, many of the principles and notions observed in Univariate KDE can be applied to the multivariate scenario. However, it is important to note that certain aspects of these concepts pose significant technical complexities in the multivariate context.
In our case, we rely on MKDE which is a non-parametric technique and, thus, it can easily and smoothly estimate the density of the 2-dimensional data space ($y^{l}, y^{h}$ values for each dimension) from point-based data
and approximate the region/space around the available points. Based on the MKDE we can have the overall picture of the structure of the data that depict the requests for processing, thus, the data required to complete these processing activities. 

\subsection{Machine Learning Based Data Selection}
\label{ocsvm}
For learning the `normal' data ($y^{l}, y^{h}$ values), exclude the outliers and combine the results with the statistically oriented analysis as presented above, we rely on the OCSVM model. To approximate as much as possible the available pairs, we adopt a non linear kernel. 
Let $\mathbf{y} = \left\lbrace \left[ y^{l}_{j}, y^{h}_{j} \right] \right\rbrace$ denote a request boundary 2-dimensional point on the local data requests described in Section \ref{overview} for the $j$th dimension. Let also the set $\mathcal{Y} = \{\mathbf{y}_{1}, \ldots, \mathbf{y}_{|\mathcal{Y}|}\}$ correspond to the boundaries points of the requests issued to node $n_{k}$. 
OCSVM \cite{SVM_for_ND} separates all $|\mathcal{Y}|$ boundary points from the origin in a feature space $\phi(\mathbf{y}) \in \mathcal{F}$ that maximizes the distance of this hyperplane to the origin. The target is to learn a decision function $F(\mathbf{y}) \in \{1, -1\}$, which captures regions of boundary points with high probability density, a.k.a. \textit{inliers}. 
$F(\mathbf{y}) = 1$ in a region captures inlier boundary points, while $F(\mathbf{y}) = -1$ captures outlier boundary points. The objective is the minimization of:  
\begin{eqnarray}
    \min_{\mathbf{y}, \xi_{i}, \rho} \frac{1}{2}\lVert \mathbf{y} \rVert^{2} + \frac{1}{\nu |\mathcal{Y}|}\sum_{i=1}^{|\mathcal{Y}|}\xi_{i}-\rho \\
    \mbox{s.t. }\phi(\mathbf{y})^{\top}\phi(\mathbf{y}_{i}) \geq \rho-\xi_{i}, \xi_{i} \geq 0, \forall i = 1, \ldots, |\mathcal{Y}|,
\end{eqnarray}
where $\nu$ indicates the upper bound on the fraction of non-inliers (boundary points regarded as `out-of-class') and the lower bound on the number of boundary points referred to as boundary Support Vectors (SVs). Then, the learnt function is:
\begin{equation}
F(\mathbf{y}) = \sum_{i=1}^{m}a_{i}K(\mathbf{y}_{i},\mathbf{y})-\rho,
\label{eq:F}
\end{equation}
with $m < |\mathcal{Y}|$ SVs, $\mathbf{y}_{1}, \ldots, \mathbf{y}_{m}$, coefficients $a_{i} > 0$ and $K(\mathbf{y}, \mathbf{y}_{i}) = \langle \phi(\mathbf{y}), \phi(\mathbf{y}_{i}) \rangle$ is a Kernel function, e.g., the Radial Basis Function $K(\mathbf{y}, \mathbf{y}_{i}) = \exp(-\frac{1}{2\sigma^{2}}\lVert \mathbf{y}-\mathbf{y}_{i} \rVert^{2})$. A negative value of $F(\mathbf{y})$ indicates that the boundary point $\mathbf{y}$ is an outlier, thus, not being included in the data filter boundary determination. Hence, the core boundary points to be considered further are:
\begin{eqnarray}
    \{\mathbf{y} \in \mathcal{Y}: F(\mathbf{y}) = 1\}.
\end{eqnarray}
Since we cope with streaming data and requests, concepts cannot be static, thus, as it is infeasible to store all data on the node due to limited resources. We, then, adopt a sliding window approach and focus on the most recent observations and requests. 
Specifically, assuming a discrete time domain $t \in \mathbb{T} = \{1, 2, \ldots \}$, a sliding window with size $W$ is defined as an ordered set of the boundary points ranked by their time indices (time index of the corresponding request), i.e., $\mathcal{W}_{t} = \{\mathbf{y}_{t-W+1}, \ldots, \mathbf{y}_{t}\}$ corresponding to requests $\{\mathcal{T}_{t-W+1}, \ldots, \mathcal{T}_{t}\}$. Based on the sliding window structure, the new boundary point at $t+1$ will be added in the window discarding the oldest point $\mathbf{y}_{t-W+1}$, i.e., $\mathcal{W}_{t+1} = \{\mathcal{W}_{t} \setminus \{\mathbf{y}_{t-W+1}\}\} \cup \{\mathbf{y}_{t+1}\}$. 

Node $n_{k}$ locally trains an OCSVM model for each dimension over the sliding windows in light of identifying the 
boundary SVs $m_{t} < W$. The amount $m_{t}$ of boundary SVs affects the boundary shape and therefore model's outlier detection capacity; theoretically, the expectation $\mathbb{E}[m]$ is controlled by the parameter $\nu$. Node $n_{k}$, based on the outcome of the OCSVM models, which eliminate the outliers (boundary points with $F(\mathbf{y}) = -1$), obtains the lowest and highest request boundaries 
$\left\lbrace \left[ \overline{y}^{l}_{j}, \overline{y}^{h}_{j} \right] \right\rbrace$, $\forall j$.

\subsection{The Proposed Data Migration Model}
\begin{figure}[h]
\centering
\includegraphics[width=0.45\textwidth]{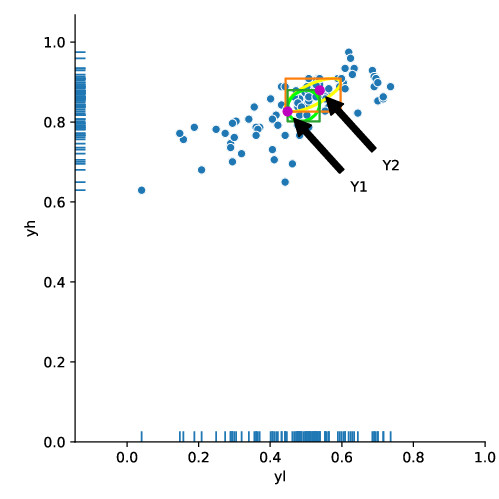}
\caption{Ensemble Model Example}
\label{fig_exm}
\end{figure}

By integrating the various components of the puzzle, we will explicate the manner in which the proposed model induces the dominant data interval that aligns with the requests of the offloaded tasks. 
Upon completion of $L_r$ requests ($L_r$ being an integer divisor of $W$), the receptor node implements to at most $W$ enqueued requests 
\begin{itemize}
    \item the MKDE, establishing the quantity of contour levels (specifically, one) and the minimum iso-proportion at which a contour line is depicted ($thresh$); and 
    \item the OCSVM model, setting an upper limit on the proportion of errors in the training set and a lower limit on the proportion of support vectors ($nu$).
\end{itemize} 
In this way, some `islands' are extracted which corresponds to the most popular data regions. To attain enhanced generalization of our model, we exploit the bounding boxes of the curves generated in the preceding stage. 
These bounding boxes are distinctly indicated in the visual representation depicted in Figure \ref{fig_exm}, where they are colored in green (left-low) and orange (right-up), respectively. 
Subsequently, the intersection (a conjunctive scheme) of the aforementioned boxes is estimated and the lower left and upper right points are determined (depicted in the Figure by dots). 
These points correspond to the Y1: $\left(y_{1}^{l},y_{1}^{h}\right)$ and Y2: $\left(y_{2}^{l},y_{2}^{h}\right)$ points depicted in the Figure. The dominant data interval is defined by the combination of the $y_{1}^{l}$ and $y_{2}^{h}$ coordinates, resulting in the largest potential range. 
Finally, the estimated data interval is migrated to those requestors which offload tasks with a frequency greater than a threshold $\omega$. This is because with our model we try to avoid overloading the network with data transfers when those refer in activities that are not popular. 
The following Algorithm presents the proposed processing for an individual dimension. The same processing stands true for the remaining dimensions and the combination of the results will formulate the final data sub-spaces that will be pushed to the requestor. 

\begin{algorithm}[hbt!]
\caption{Proactive Data Migration}\label{alg:one}
\begin{algorithmic}
\FORALL{$\mathcal{T} =\left\lbrace \left[ y^{l}, y^{h} \right] \right\rbrace$}
    \IF{$\left[ y^{l}, y^{h} \right] \in \mathbf{x}$}
        \STATE Execute $\mathcal{T}$ at $n_i$
        \IF{$Requestor$ $n_j$ $\in \mathcal{N}$}
           \STATE Add $\left[ y^{l}, y^{h} \right]$ in $Qr_i$ \COMMENT{Requests Queue}
            \STATE Update $Fr_{ij}$\COMMENT{Frequency Map}
            \IF{$|Qr_i| \mod{L_r} = 0$}
                \STATE Deploy MKDE and OCSVM to Infer $\left[ \hat y^{l}, \hat y^{h} \right]$
                \STATE Migrate $\left[ \hat y^{l}, \hat y^{h} \right]$ to $Requestors$ $n_k$, $\sum Fr_{ik} \geq \omega$
            \ENDIF
        \ENDIF
    \ELSE
        \STATE Offload $\mathcal{T}$ to $n_d$ \COMMENT{$n_d$ peer node in the same cluster}
    \ENDIF
\ENDFOR
\end{algorithmic}
\end{algorithm}

\section{Experimental Evaluation}
\label{setup}

\subsection{Performance Metrics, Datasets \& Setup}
We provide a comprehensive analysis of the performance of the suggested model (i.e., \textit{Model}) in order to reveal its inherent advantages and limitations. Specifically, we analyze our model in relation to two alternative approaches with respect to how a model infers the dominant data intervals to be migrated, chooses the peer nodes where they will be located, and subsequently determines the temporal occurrences of these actions:
\begin{itemize}
  \item the Random Data Migration (\textit{Random-DM}) model randomly selects the dominant data intervals to be migrated to a random subset of peer nodes without considering any contextual information; and 
  \item the DBSCAN Data Migration (\textit{DBSCAN-DM}) model deduces the dominant pairs to be migrated to peer nodes which submit data requests more frequently, based on DBSCAN algorithm. The dominant data pair corresponds to the `centroid' of the expanded cluster with the highest frequency.
\end{itemize}
The \textit{Random-DM} is chosen for the purpose of simulating a mutable patternless way of choosing the data interval to migrate and the nodes to host. On the other hand, the \textit{DBSCAN-DM} model is selected to simulate an approach based on utility, in which the most frequently requested data interval is migrated or replicated to PEC nodes exhibiting the strongest interest in that. 
The likelihood of making a migration when adopting the \textit{Random-DM} model is derived from the (average) proportion between the total number of requests that initiate a migration and the overall number of requests throughout an experimental process of \textit{Model}. 
Whereas, when the \textit{DBSCAN-DM} model is implemented, the estimation of the data intervals and their consequent migrations are carried out by fulfilling $L_r$ requests. In both cases, the quantity of nodes in which the migrations (\textit{qnm}) take place is determined by the proportion of the overall number of migrations to the total number of requests that are accompanied by migrations, as they are derived from the experimental adoption of the \textit{Model}. 
In addition, the assessment of the alternative approaches is conducted in a highly dynamic setting, wherein mobile users of the pervasive applications have the ability to continuously submit requests at varying rates (\textit{req\_step}).

In order to evaluate the ability of nodes to infer the prevailing data interval and migrate it to the appropriate peer nodes that become capable of handling upcoming requests directed towards them in the foreseeable future, we adopt the following metric:
\begin{equation}\label{metric_μ}
\mu =\frac{TP}{TP+FP+UnM} 
\end{equation}
in two different versions, i.e., $\mu_s$ (strict $\mu$) and $\mu_r$ (relaxed $\mu$), respectively. $TP$ refers to a True Positive classification - the data migration is correctly chosen when forthcoming requests can be satisfied at the new host PEC node, that is to say, the essential data pertaining to the requests are encompassed within the updated available data interval. In the situation of the $\mu_s$, we analyze the request that immediately follows to characterize the classification, whereas in the situation of the $\mu_r$, we study the subsequent requests across the span of window $L_r$. 
$FP$ corresponds to a False Positive classification - data migration is incorrectly decided when the recently data interval migrated fails to meet the next request or the window $L_r$ of requests, respectively. 
$UnM$ stands for unnecessary migrations that take place when no request intervenes between two successive migration operations on a given requestor. It is worth mentioning that when utilizing $L_r$ as a reference frame, the number of hits within it is documented ($\kappa$).
In addition, if we adopt the definition of the distance between two intervals as the Euclidean distance between the 2-dimensional points corresponding to their endpoints \cite{Kosheleva}, we can conclude to the following metric:
\begin{equation}\label{metric_δ}
\delta =\frac{1}{r_N} \sum_{i=1}^{N} \sum_{j=1}^{{\tilde{r}_i}} (Y_{ij}-\hat{Y_i})^2
\end{equation}
where $r_{N}$ corresponds to the total number of requests submitted to the PEC nodes ecosystem, ${\tilde{r}_i}$ refers to the number of data requests that are not available at the corresponding node $n_i$, and $Y_{ij}$ and $\hat{Y_i}$ are the 2-dimensional points of an unserviceable request and the $n_i$'s available data interval, respectively. The metric $\delta$ enables us to obtain a sense of the extent to which the intervals of the inferred/migrated available data deviate from the unsatisfiable requests.

The assessment of the effectiveness of the proposed scheme is executed by leveraging data from real-life scenarios. In order to formulate the normalized endpoints of the data requests, we used the lowest and highest observed temperatures available in the `Comparative Climatic Data' tables of NOAA\footnote{https://www.ncei.noaa.gov/products/land-based-station/comparative-climatic-data}. Another dataset that has been selected is the `City Subway Stations' from NYC Open Data\footnote{https://data.cityofnewyork.us/Transportation/subway-stations/jaej-er89}. The dataset is provided by the Metropolitan Transportation Authority and contains various cognitive elements such as name and coordinates. 
We make the assumption that a certain number of nodes, $N\in\{20,50\}$, are randomly distributed among a subset of the subway station locations. Following this, K-means clustering is applied to these nodes. In order to determine the optimal number of clusters, we make use of the elbow method as well as silhouette analysis or score techniques. For the purposes of experimentation, we consider $|C|\in\{3, 4\}$ ($|C|$ is the number of clusters), which is also the cardinality of sinks, one for each cluster. The nodes’ allocation and clustering in NYC are presented in \cite{boulkol}. 
Finally, we adopt a sample of FOILing NYC's Taxi Trip dataset of Chris Wong\footnote{https://chriswhong.com/open-data/foil\_nyc\_taxi/} to facilitate users and their IoT devices in navigating the streets of NYC ($trips\in \{1000,2000,3000\}$) and submitting data requests at regular steps of miles, $req\_step\in\{0.5,1,2\}$. The performance outcomes are extracted by setting $thresh=0.8$ and $nu=0.8$ when adopting MKDE and OCSVM methods, respectively in our proposed ensemble model and $eps=0.0296$ and $min\_samples=4$ in case of \textit{DBSCAN-DM} model. In both cases $L_r=20$ and $W=5*L_r$.

Our custom simulator is implemented in Python. A class is dedicated to imitate the existing nodes in PEC ecosystem identified by a large number of parameters such as id, geographic location, list of peers in same cluster, endpoints of available data intervals and so on. In addition, the main class employed for the execution of our simulations establishes a collection of nodes and available data intervals and carries out a high number of iterations ($trips$) as indicated by the previously described datasets and the alternative approaches for data migration. The nodes' initial available data intervals are set using particular random seeds in order to establish a common framework for comparison. Subsequently, the decision-making process is motivated by data requests and mobility of IoT devices that arise from real datasets, with a range of experimental scenarios being adopted. These scenarios involve distinct schemes and different combinations of the aforementioned values.

\subsection{Performance Assessment}
\label{assessment}
In this subsection, we present a thorough analysis of the previously indicated performance metrics with the aim of providing the complete understanding of the conducted experiments. First, we present findings concerning the nodes' behavior with regard to the migration of data intervals, whereby the suggested methodology is implemented. As demonstrated in Table \ref{table:2}, the migration of data takes place in approximately 2.01\% and 4.84\% of total requests in ecosystem when 20 or 50 nodes, respectively, participate in it. In fact, approximately 1.978 and 4.262 on average migration actions occurs in such instances (Table \ref{table:3}). Hence, these statistics are duly acknowledged in the process of adopting alternative methodologies in order to establish a unified basis for comparison. The behavior of the proposed model is inherently influenced by the quantity of nodes and requests. As $N$ and the number of requests for each trip (inversely proportional to $req\_step$) increase, so does the number of migration actions. This phenomenon occurs due to the increased and dispersed need for the accomplishment of tasks. 

\begin{table}[h!]
\centering
\label{migration_percentage}
\caption{Percentage of Data Request with Migration}
\begin{tabular}{cccccc}
\hline
\hline
\multicolumn{1}{l}{}    & \multicolumn{1}{l}{\multirow{2}{*}{$trips$}} & \multicolumn{3}{c}{$req\_step$} & \multicolumn{1}{l}{\multirow{2}{*}{\textbf{Average}}} \\
\multicolumn{1}{l}{}    & \multicolumn{1}{l}{}                       & 2.0        & 1.0        & 0.5     & \multicolumn{1}{l}{}                         \\
\hline
\multirow{3}{*}{$N = 20$} & 1000                                       & 2.03\%   & 2.42\%   & 2.57\%  & \multirow{3}{*}{\textbf{2.01\%}}             \\
                        & 2000                                       & 1.59\%   & 1.70\%   & 2.04\%  &                                              \\
                        & 3000                                       & 1.64\%   & 1.94\%   & 2.20\%  &                                              \\
\hline
\multirow{3}{*}{$N = 50$} & 1000                                       & 5.57\%   & 5.82\%   & 5.90\%  & \multirow{3}{*}{\textbf{4.84\%}}             \\
                        & 2000                                       & 4.29\%   & 4.36\%   & 4.64\%  &                                              \\
                        & 3000                                       & 4.10\%   & 4.34\%   & 4.50\%  &                                              \\
\hline
\hline
\end{tabular}
\label{table:2}
\end{table}

\begin{table}[h!]
\centering
\label{migrations_number}
\caption{Number of Data Migration Actions}
\begin{tabular}{cccccc}
\hline
\hline
\multicolumn{1}{l}{}    & \multicolumn{1}{l}{\multirow{2}{*}{$trips$}} & \multicolumn{3}{c}{$req\_step$} & \multicolumn{1}{l}{\multirow{2}{*}{\textbf{Average}}} \\
\multicolumn{1}{l}{}    & \multicolumn{1}{l}{}                       & 2.0        & 1.0        & 0.5     & \multicolumn{1}{l}{}                         \\
\hline
\multirow{3}{*}{$N = 20$} & 1000                                       & 1.889    & 1.926    & 2.040   & \multirow{3}{*}{\textbf{1.978}}              \\
                        & 2000                                       & 1.867    & 1.940    & 2.106   &                                              \\
                        & 3000                                       & 1.898    & 1.962    & 2.178   &                                              \\
\hline
\multirow{3}{*}{$N = 50$} & 1000                                       & 3.970    & 4.259    & 4.188   & \multirow{3}{*}{\textbf{4.262}}              \\
                        & 2000                                       & 4.356    & 4.255    & 4.190   &                                              \\
                        & 3000                                       & 4.388    & 4.432    & 4.324   &                                              \\
\hline
\hline
\end{tabular}
\label{table:3}
\end{table}

\begin{figure*}%
\centering
\begin{subfigure}{.95\textwidth}
\includegraphics[width=0.9\textwidth]{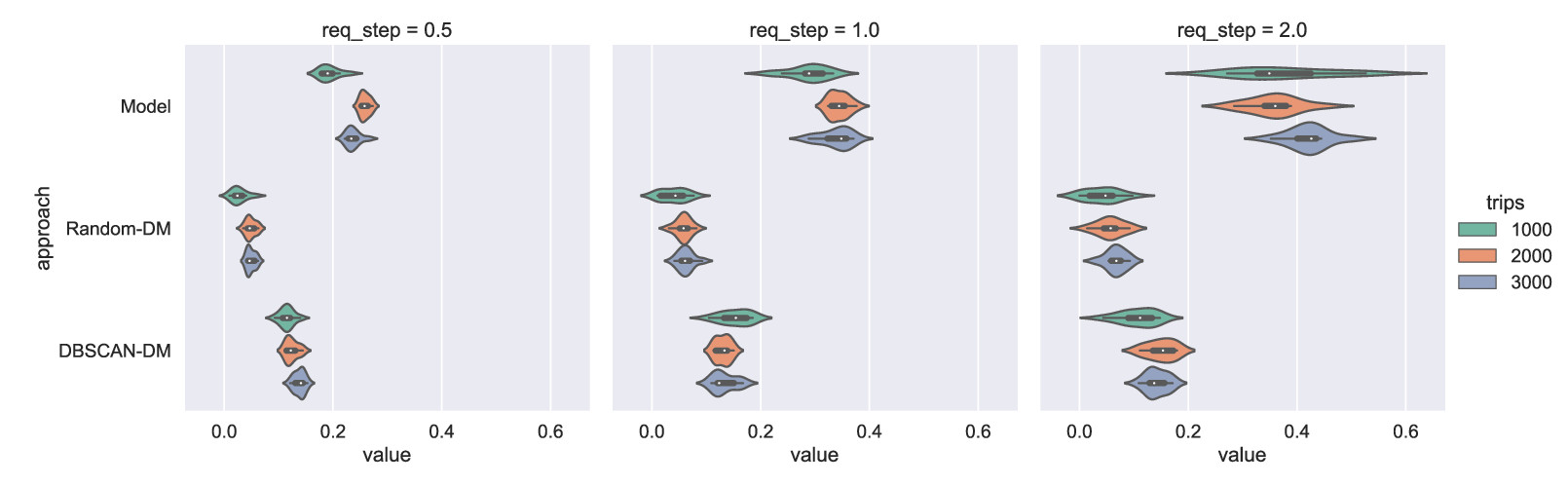}%
\caption{$N = 20$}
\label{fig_sm20}
\end{subfigure}\hfill%
\begin{subfigure}{.95\textwidth}
\includegraphics[width=0.9\textwidth]{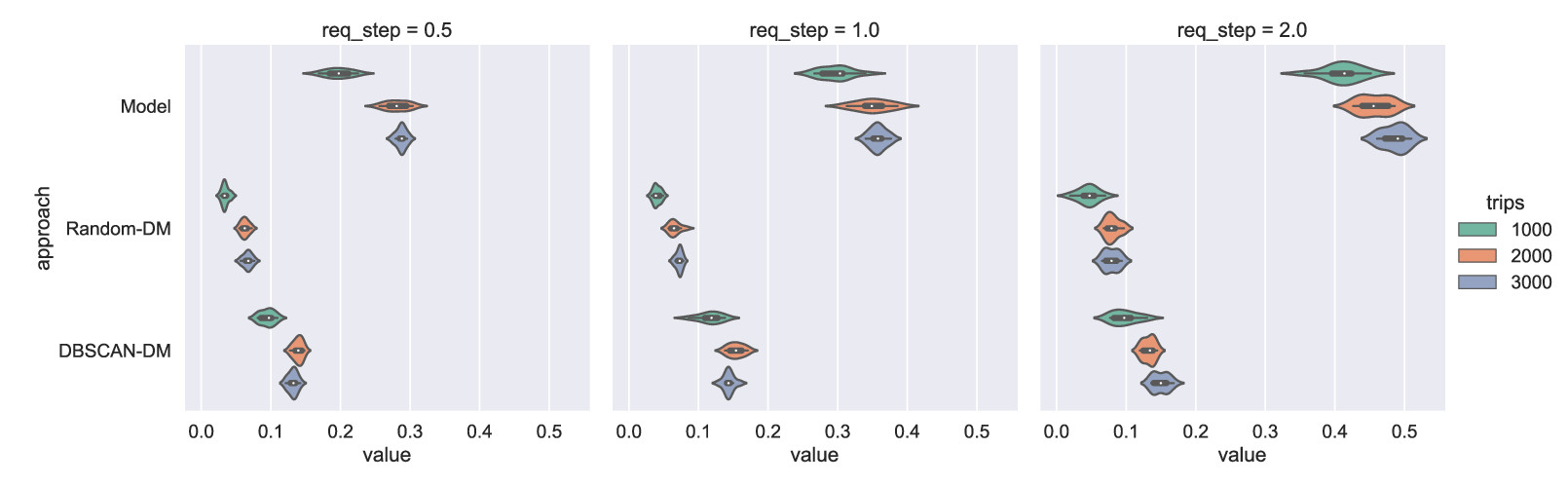}%
\caption{$N = 50$}
\label{fig_sm50}
\end{subfigure}%
\caption{Strict Performance Assessment - $\mu_s$}
\label{fig_sm}
\end{figure*}

\begin{figure*}%
\centering
\begin{subfigure}{.95\textwidth}
\includegraphics[width=0.9\textwidth]{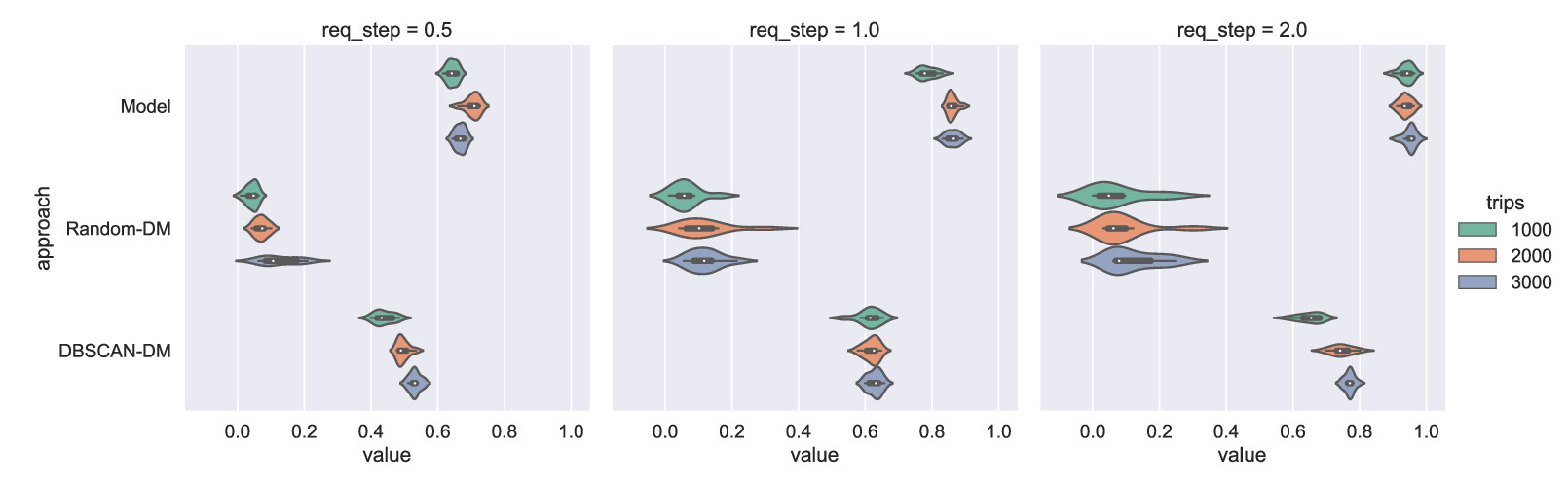}%
\caption{$N = 20$}
\label{fig_rm20}
\end{subfigure}\hfill%
\begin{subfigure}{.95\textwidth}
\includegraphics[width=0.9\textwidth]{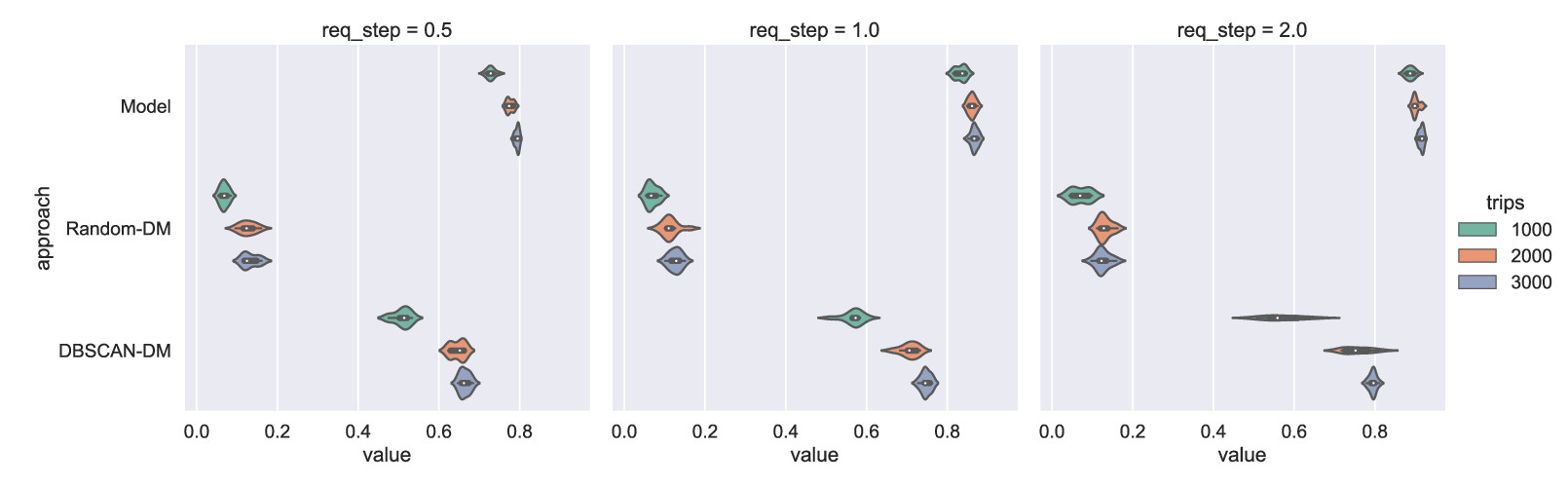}%
\caption{$N = 50$}
\label{fig_rm50}
\end{subfigure}%
\caption{Relaxed Performance Assessment - $\mu_r$}
\label{fig_rm}
\end{figure*}

Figures \ref{fig_sm} and \ref{fig_rm} demonstrate the $\mu_s$ and $\mu_r$ outcomes when $N\in\{20,50\}$. As anticipated, the alternative methodologies exhibit superior performance when the more relaxed metric calculation method is adopted - $\mu_r$, regardless of the number of nodes. This conclusion can be attributed to the substantial growth in the amount of $TP$, juxtaposed with a conspicuous decline in the number of $FP$ classifications. Upon initial observation, the $Model$ displays comparable behavior whether 20 or 50 nodes participates within ecosystem. In both modes of performance evaluation, the $Model$ demonstrates better outcomes as the $req\_step$ size increases. The increase in $req\_step$ size is accompanied by a decrease in the overall number of requests during a complete trip, as well as a restriction in their dispersion. This restriction enables the existing infrastructure to respond more efficiently and effectively in meeting their satisfaction. It is a significant characteristic that when the metric $\mu_r$ is adopted and the $req\_step=2.0$, the value of the metric approaches unity across all three distinct values of $trips$, while also accounting for variations in the cardinality of $N$. 

Another aspect that can be highlighted is the reduction in the size of the proposed model violins when applying the $\mu_r$ metric, suggesting a constrained interquartile range of the recorded values. If an attempt is made to perform a comparative evaluation, it becomes evident that the proposed model consistently outperforms the remaining schemas across all experimental scenarios implemented and regardless of the $\mu$ metric employed. In the case of adopting the metric $\mu_s$, the ensemble $Model$ accomplishes values that are two, three, and on certain occasions, four times superior to those of $DBSCAN-DM$. The disparity becomes even more significant when the values of the model $Random-DM$ are utilized as a benchmark. The outcomes can guide us to the secure deduction that the $Model$ contributes to more efficient decision-making of the PEC nodes associated with the determination of the appropriate data interval to be migrated, as well as the selection of the neighboring nodes it is necessary to carry it out, thus enabling them to effectively meet, in a collaborative manner, the demands of the dynamic environment in which they elaborate. 

\begin{figure}[H]
\centering
\begin{subfigure}{0.48\textwidth}
    \centering
    \includegraphics[width=1\linewidth]{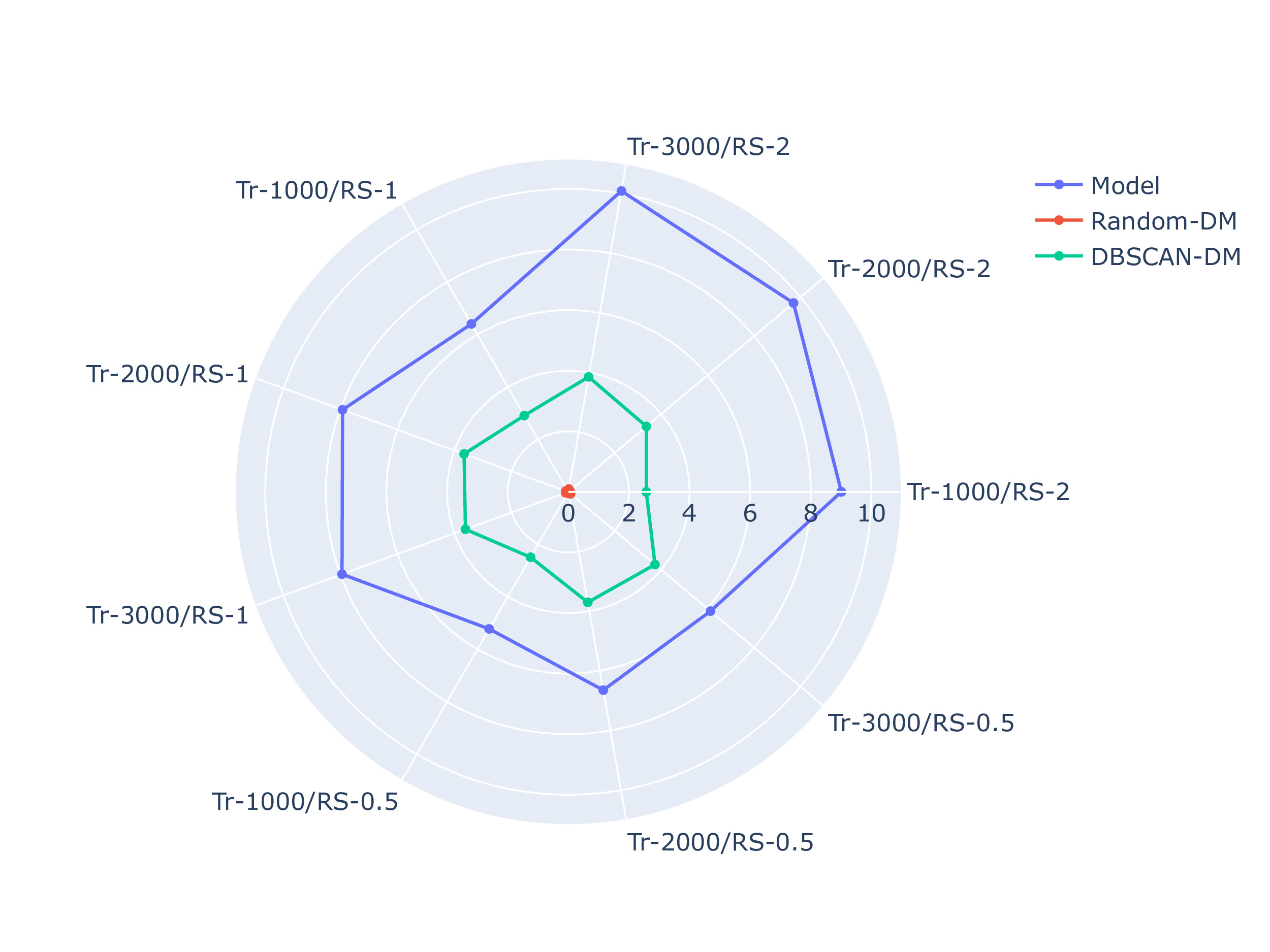}  
    \caption{$N = 20$}
\end{subfigure}
\begin{subfigure}{0.48\textwidth}
    \centering
    \includegraphics[width=1\linewidth]{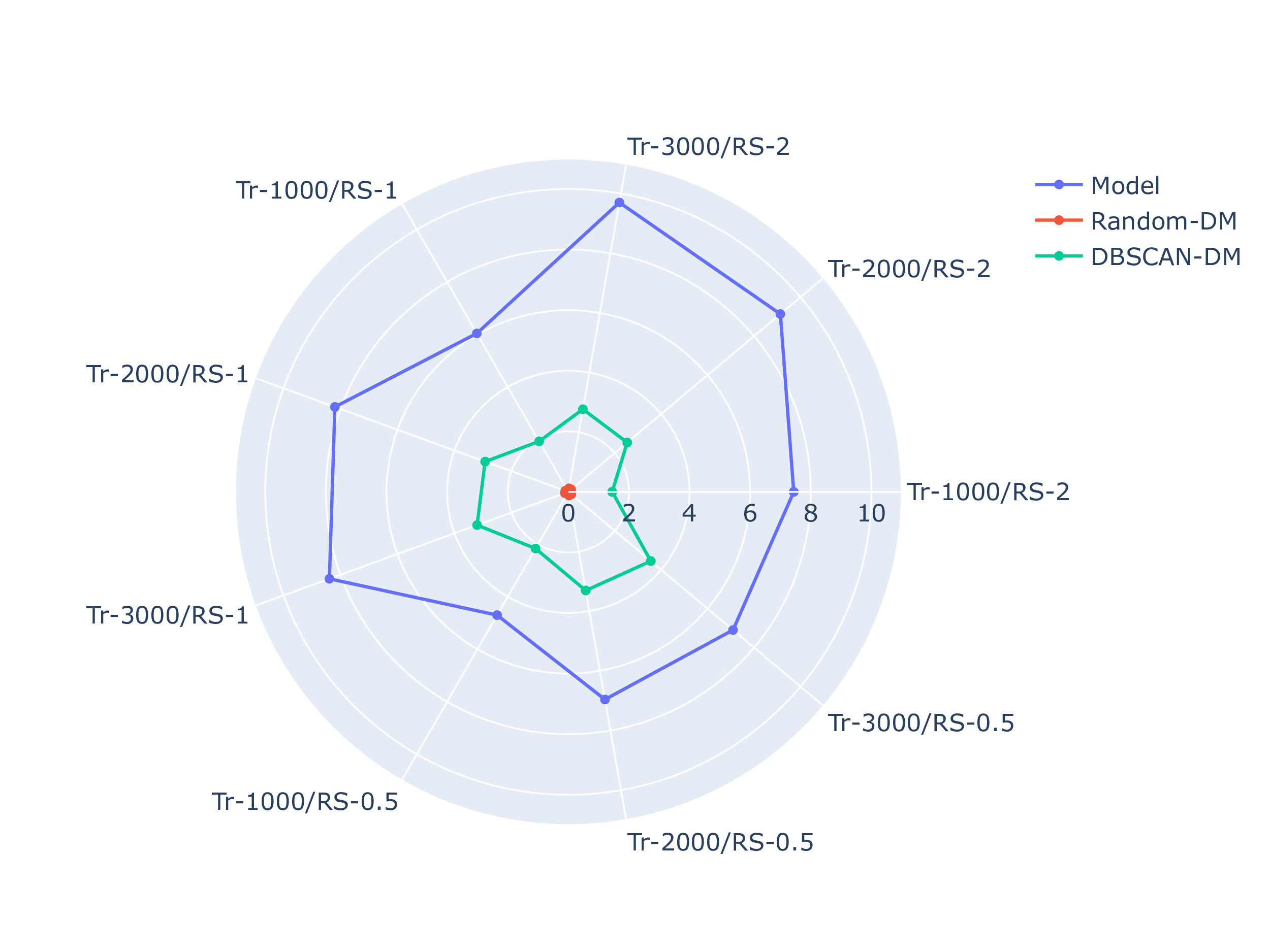}  
    \caption{$N = 50$}
\end{subfigure}
\caption{Average Number of Hits within $L_r$ - $\kappa$}
\label{fig_avh}
\end{figure}

\begin{figure}[H]
\centering
\begin{subfigure}{0.48\textwidth}
    \centering
    \includegraphics[width=1\linewidth]{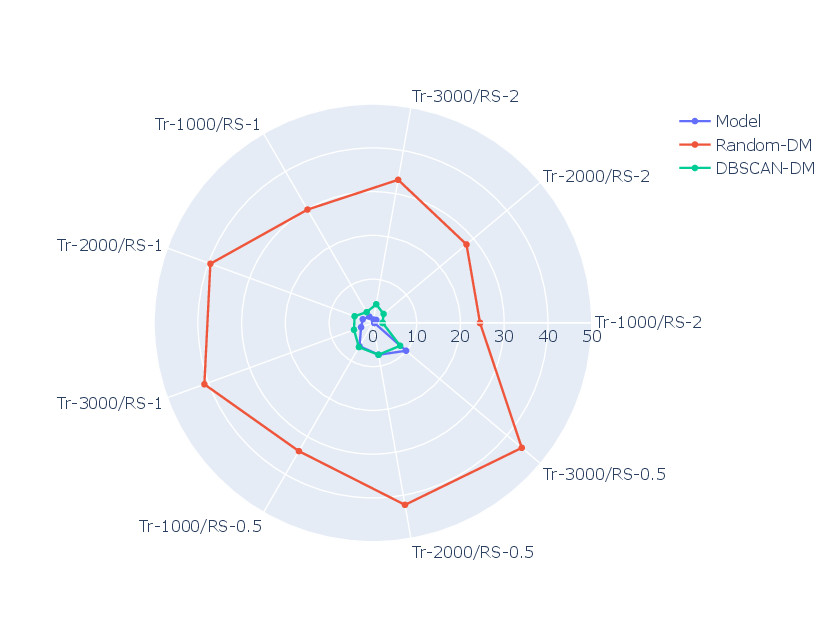}  
    \caption{$N = 20$}
\end{subfigure}
\begin{subfigure}{0.48\textwidth}
    \centering
    \includegraphics[width=1\linewidth]{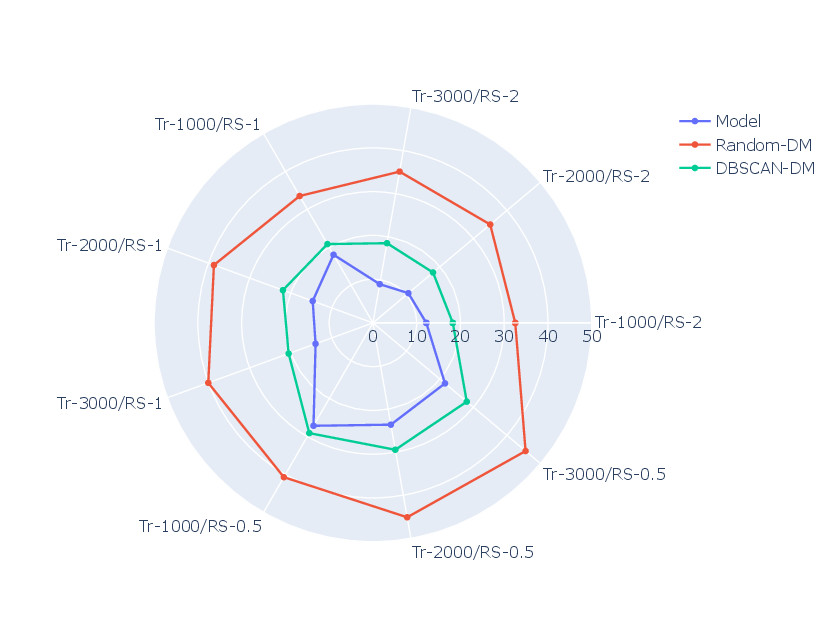}  
    \caption{$N = 50$}
\end{subfigure}
\caption{Unnecessary Migrations Ratio}
\label{fig_unm}
\end{figure}

The findings of the models under investigation depicted in Figures \ref{fig_avh} and \ref{fig_unm} additionally contribute to the preceding conclusion. In the first one, we are able to observe the frequency at which the data interval migration to a node enables it to fulfill forthcoming requests made by users or IoT devices. The $Model$'s $\kappa$ values exhibit greater magnitudes across all expanded instances. The minimum average value of the $Model$ for $\kappa$, with 20 nodes interacting, is approximately 5.22, whereas the maximum value is almost 10. When referring to 50 nodes, the values range between 4.7 and 9.7. For 20 nodes, the corresponding values of $DBSCAN-DM$ are 2.49 and 3.85, falling within the interval [1.45, 3.55] in the case of 50 nodes. These values are significantly lower than unity when adopting baseline $Random-DM$. A novel aspect of the nodes' behavior, with regards to their efficacy in decision-making, is uncovered in Figure \ref{fig_unm}. This pertains to the quantity of superfluous migrations conducted by the nodes, which indirectly illuminates the models' performance in terms of time, specifically when they opt to execute the data migrations. The interesting point in this figure is that model $DBSCAN-DM$ exhibits similar behavior to the proposed model. The polygons are almost identical when $N=20$, and the curves display minimal fluctuation at $N=50$. Both models decide in a comparable way when to migrate the induced data interval to the interested peer nodes. 

\begin{figure}[H]
\centering
\begin{subfigure}{0.48\textwidth}
    \centering
    \includegraphics[width=1\linewidth]{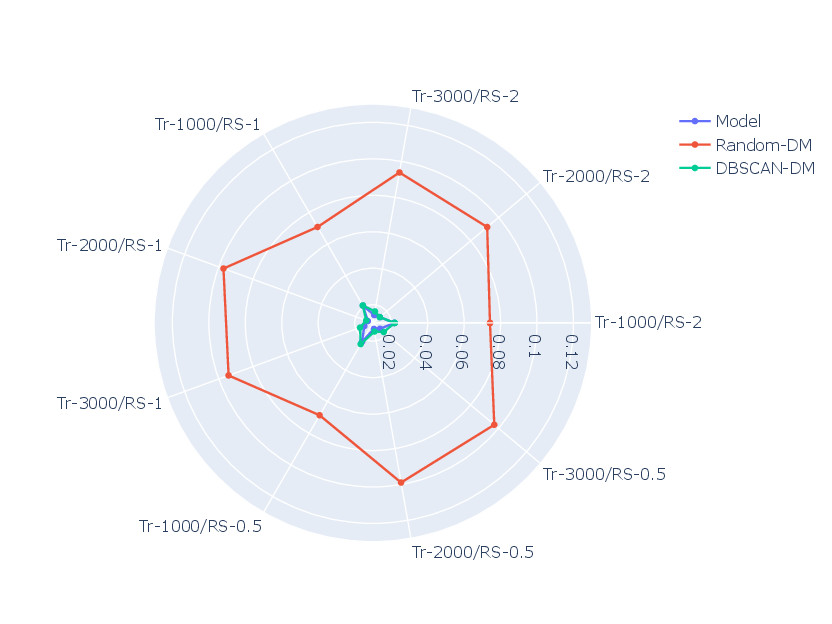}  
    \caption{$N = 20$}
\end{subfigure}
\begin{subfigure}{0.48\textwidth}
    \centering
    \includegraphics[width=1\linewidth]{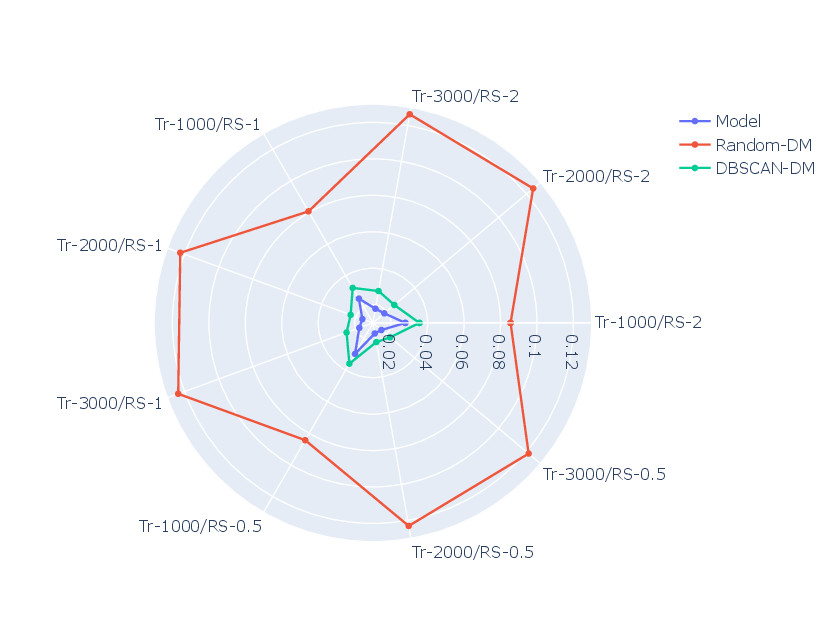}  
    \caption{$N = 50$}
\end{subfigure}
\caption{Distance of Unsatisfiable Requests - $\delta$}
\label{fig_delta}
\end{figure}

Finally, in Figure \ref{fig_delta}, we are able to get an idea of the Euclidean distance between the 2D points that correspond to the unfulfilled requests from the nearest node and the data intervals available to them. It seems that all the various methodologies produce a greater mean distance when $N=50$ as opposed to when $N=20$. In any case when baseline $Random-DM$ is adopted, the requests that cannot be satisfied extend further beyond the endpoints of the data intervals currently accessible to the nodes that already exist. The distances corresponding to the two remaining methods are comparatively smaller, with slightly better results, when model $Model$ is employed.

\section{Conclusion}
\label{conclusions}
Various challenges should be addressed in PEC environments, particularly concerning the management of data and the accessible services, while considering the constrained computational capabilities of the devices/nodes participating in them. In this paper, we present a novel ensemble scheme incorporating MKDE scheme and an OCSVM model, enabling the PEC nodes to identify the data intervals that exert significant influence on the requests streams of the offloaded tasks. The comparative assessment conducted on actual datasets demonstrates that the suggested methodology has the ability to provide effective decision-making. It enables nodes to accurately deduce the dominant data interval (\textbf{what}), the potential requestors (\textbf{where}), and indirectly determine the suitable timing (\textbf{when}) to accommodate the fluctuations in demand within dynamic ecosystem.
Our forthcoming research plans encompass the implementation of federated learning framework that capitalizes on the suggested ensemble model, as well as the assessment of its performance in comparison to diverse distributed clustering models.


\section{Biography Section}
\vskip -2\baselineskip plus -1fil
\begin{IEEEbiography}
[{\includegraphics[width=1in,height=1.25in,clip,keepaspectratio]{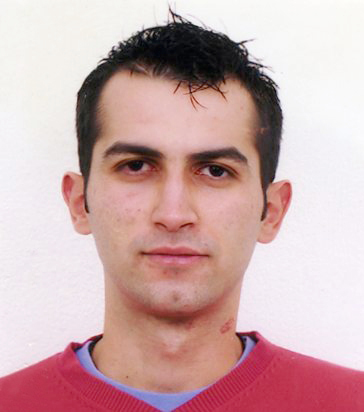}}]
{Georgios Boulougaris}
Georgios Boulougaris is an Informatics Teacher in Secondary Education. He received a B.Sc. in Applied Informatics from the Department of Applied Informatics at the University of Macedonia, a M.Sc. in Advanced Information Systems from the Department of Informatics and Telecommunications at the National and Kapodistrian University of Athens and a M.Ed in Studies in Education from the Hellenic Open University. Currently, he is a Ph.D. candidate in the Department of Informatics and Telecommunications, University of Thessaly and he is a member of the Intelligent Pervasive Systems (iPRISM) research group (http://www.iprism.eu). His research interests include Intelligent Systems, Distributed Systems, Pervasive Computing, Machine Learning, Internet of Things, Edge Computing and Computer Science Pedagogy.
\end{IEEEbiography}
\vskip -2\baselineskip plus -1fil
\begin{IEEEbiography}
[{\includegraphics[width=1in,height=1.25in,clip,keepaspectratio]{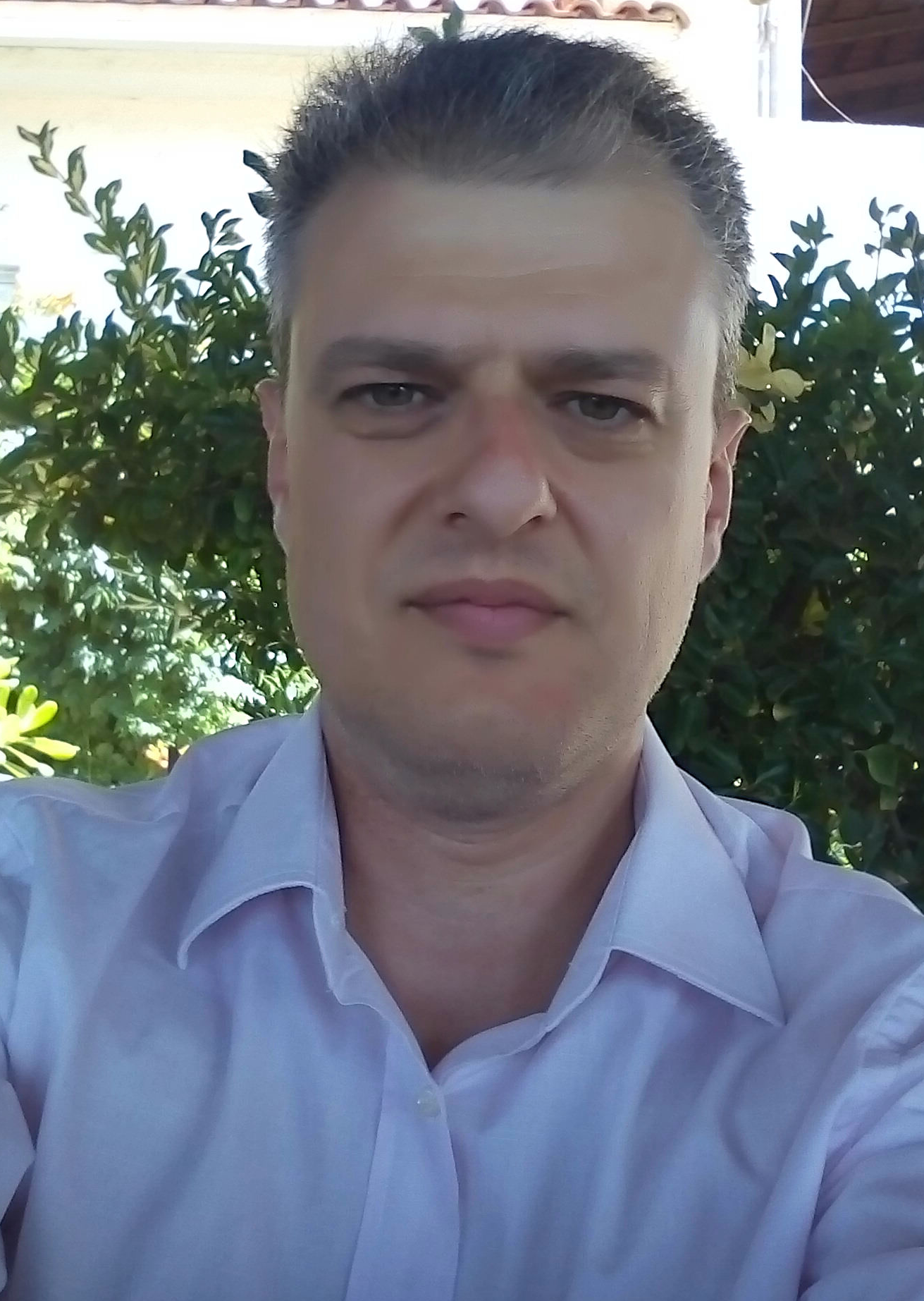}}]
{Kostas Kolomvatsos}
Dr Kostas Kolomvatsos received his B.Sc. in Informatics from the Department of Informatics at the Athens University of Economics and Business, his M.Sc. and his Ph.D. in Computer Science from the Department of Informatics and Telecommunications at the National and Kapodistrian University of Athens. Currently, he serves as an Assistant Professor in the Department of Informatics and Telecommunications, University of Thessaly. He was a Marie Skłodowska Curie Fellow (Individual Fellowship) at the School of Computing Science, University of Glasgow. His research interests are in the definition of Intelligent Systems adopting Machine Learning, Computational Intelligence and Soft Computing for Pervasive Computing, Distributed Systems, Internet of Things, Edge Computing and the management of Large Scale Data. He is the author of over 130 publications in the aforementioned areas.
\end{IEEEbiography}

\begin{thebibliography}{1}

\bibitem{alam}
Alam, M. G. R., Tun, Y. K., \& Hong, C. S., 'Multi-agent and reinforcement learning based code offloading in mobile fog', in ICOIN, 2016, 285-–290.



\bibitem{ref2}
Awadalla, M., 'Task Mapping and Scheduling in Wireless Sensor Networks', International Journal of Computer Science, 440(4), 2013.


\bibitem{ref32}
Bhardwaj, K., Agrawal, P., Gavrilovska, A., \& Schwan, K., 'AppSachet: Distributed App Delivery from the Edge Cloud', 7th Intl. Conf. Mobile Computing, Applications, and Services, 2015, 89--106.



\bibitem{bellavista2}
Bellavista, P., Zanni, A., \& Solimando, M., 'A migration-enhanced edge computing support for mobile devices in hostile environments', 13th IWCMC, 2017, 957–-962.

\bibitem{Breitbach}
Breitbach, M., Schäfer, D., Edinger, J., \& Becker, C., 'Context-Aware Data and Task Placement in Edge Computing Environments', IEEE International Conference on Pervasive Computing and Communications (PerCom), 2019.

\bibitem{boulougaris}
Boulougaris, G., Kolomvatsos, K., ‘A QoS-aware, Proactive Tasks Offloading Model for Pervasive Applications’, in 9th International Conference on Future Internet of Things and Cloud (FiCloud), 22-24 Aug, Rome, Italy, 2022.

\bibitem{boulkol}
Boulougaris, G., Kolomvatsos, K., An Inference Mechanism for Proactive Service Migration at the Edge', in IEEE Transactions on Network and Service Management, vol. 20, no. 4, pp. 4505-4516, Dec. 2023.

\bibitem{coltin}
Coltin, B., Veloso, N., 'Mobile Robot Task Allocation in Hybrid Wireless Sensors Networks', in 2010 International Conference on Intelligent Robots and Systems, 2010.



\bibitem{elzeki}
Elzeki, O. M., Rashad, M. Z., \& Elsoud, M. A., 'Overview of Scheduling Tasks in Distributed Computing Systems', International Journal of Soft Computing and Engineering (IJSCE), vol. 2(3), 2012, pp. 2231--2307.



\bibitem{hardle}
Härdle, W., Müller, M., Sperlich, S., \& Werwatz, A., ‘Nonparametric and Semiparametric Models: An Introduction’, ebook, 2004, available at $http://sfb649.wiwi.hu-berlin.de/fedc_homepage/xplore/ebooks/html/spm/$

\bibitem{Hsieh}
Hsieh, L. T., Liu, H., Guo, Y., \& Gazda, R., 'Task Management for Cooperative Mobile Edge Computing', 2020 IEEE/ACM Symposium on Edge Computing (SEC), San Jose, CA, USA, 2020, pp. 352--357.

\bibitem{hu}
Hu, X., Xu, B., 'Task Allocation Mechanism Based on Genetic Algorithm in Wireless Sensor Networks', in ICAIC, 2011.

\bibitem{huang}
Huang, T., Lin, W., Li, Y., He, L., \& Peng, S., ‘A Latency-Aware Multiple Data Replicas Placement Strategy for Fog Computing’, J. Signal Process. Syst. 2019, 91, 1191–1204.






\bibitem{kolomvatsosfgcs2}
Kolomvatsos, K., ‘A proactive inference scheme for data-aware decision making in support of pervasive applications’, Future Generation Computer Systems, vol. 136, 2022, pp. 193—204. 

\bibitem{kolomvatsostnsm}
Kolomvatsos, K., Anagnostopoulos, C., 'A Proactive Statistical Model Supporting Services and Tasks Management in Pervasive Applications', IEEE Transactions on Network and Service Management, vol. 19, no. 3, pp. 3020-3031, Sept. 2022.


\bibitem{kolomvatsostkde}
Kolomvatsos, K., Anagnostopoulos, C., Koziri, M., \& Loukopoulos, T., 'Proactive \& Time-Optimized Data Synopsis Management at the Edge', IEEE TKDE, 2020.

\bibitem{kolomvatsosFGCS}
Kolomvatsos, K., Anagnostopoulos, A., 'Multi-criteria Optimal Task Allocation at the Edge', Future Generation Computer Systems, 93:358--372, 2019.


\bibitem{kolomvatsosfusion}
Kolomvatsos, K., Anagnostopoulos, C., \& Hadjiefthymiades, S., 'Data fusion and type-2 fuzzy inference in contextual data stream monitoring', IEEE TSMC:Systems, 47(8), 2016, 1839--1853. 





\bibitem{Kosheleva}
Kosheleva, O., Kreinovich, V. Euclidean Distance Between Intervals Is the Only Representation-Invariant One. 2020.

\bibitem{kumar}
Kumar, V. A., Kumar, A., Batth, R. S., Rashid, M., Gupta, S. K., \& Raghuraman, M., ‘Efficient data transfer in edge envisioned environment using artificial intelligence based edge node algorithm’, Transactions on Emerging Telecommunications Technologies, Wiley, 2021;32:e4110.

\bibitem{Kurkovsky}
Kurkovsky, S., 'Pervasive computing: Past, present and future', ITI 5th International Conference on Information and Communications Technology, Cairo, Egypt, 2007, pp. 65-71.


\bibitem{le}
Le, H. Q., Al-Shatri, H., \& Klein, A., ‘Efficient resource allocation in mobile-edge computation offloading: completion time minimization’, IEEE International Symposium on Information Theory (ISIT), Aachen, Germany, 2017:2513–2517.


\bibitem{li}
Li, Q., Wen, Z., Wu, Z., Hu, S., Wang, N., Li, Y., Liu, X., \& He, B., ‘A Survey on Federated Learning Systems: Vision, Hype and Reality for Data Privacy and Protection’, 2021, IEEE Transactions on Knowledge and Data Engineering, doi: 10.1109/TKDE.2021.3124599.

\bibitem{li2}
Li F, Wang D. 5G Network Data Migration Service Based on Edge Computing. Symmetry. 2021; 13(11):2134.




\bibitem{mosel}
Mosel, E., Vigoda, E., 'Limitations of Markov Chain Monte Carlo Algorithms for Bayesian Inference of Phylogeny', The Annals of Applied Probability, 2006, vol. 16, 4, pp. 2215--2234.


\bibitem{najam}
Hassan, N., Gillani, S., Ahmed, E., Yaqoob, I., \& Imran, M., 'The Role of Edge Computing in Internet of Things', IEEE Communications Magazine, 2018.

\bibitem{ngan}
Ngan, H. Y., Yung, N. H., \& Yeh, A. G., 'A comparative study of outlier detection for large-scale traffic data by one-class SVM and kernel density estimation', Proc. SPIE 9405, Image Processing: Machine Vision Applications VIII, 94050I, 2015.


\bibitem{Ouyang}
Ouyang, T., Chen, X., Zeng, L., \& Zhou, Z., 'Cost-Aware Edge Resource Probing for Infrastructure-Free Edge Computing: From Optimal Stopping to Layered Learning', IEEE Real-time systems, 2019.



\bibitem{Ouyang2}
Ouyang, T., Li, R., Chen, X., Zhou, Z., \& Tang, X., 'Adaptive user-managed service placement for mobile edge computing: An online learning approach', IEEE INFOCOM, 2019, pp. 1468--1476.

\bibitem{Ouyang1}
Ouyang, T., Zhou, Z., \& Chen, X., 'Follow me at the edge: Mobility-aware dynamic service placement for mobile edge computing', IEEE Journal on Selected Areas in Communications, 36(10), 2018, pp. 2333--2345, 2018.

\bibitem{rao}
Rao, G. M., Srinivas, K., Samee, S., Venkatesh, K., Dadheech, P., Raja, L., \& Yagnik, G. , ‘A Secure and Efficient Data Migration Over Cloud Computing’, IOP Conf. Ser.: Mater. Sci. Eng., 2021, 1099, 012082.


\bibitem{razavinegad}
Razavinegad, A., 'Task Allocation In Robot Mobile Wireless Sensor Networks', Int. Journal of Scientific \& Technology Research, 3(6), 2014.

\bibitem{ref26}
Sardellitti, S., Scutari, G., \& Barbarossa, S., 'Joint  Optimisation of Radio and Computational Resources for Multicell Mobile-Edge Computing', IEEE Transactions on Signal and Information Processing over Networks, 1(2):89-–103, 2015.

\bibitem{SVM_for_ND}
Schölkopf, B., Williamson, R. C., Smola, A., Shawe-Taylor, J., \& Platt, J., 'Support Vector Method for Novelty Detection' in S. Solla and T. Leen and K. M\"{u}ller, 'Advances in Neural Information Processing Systems', MIT Press, vol. 12, 1999.

\bibitem{ref34}
Simoens, P., Xiao, Y., Pillai, P., Chen, Z., Ha, K., \& Satyanarayanan, M., 'Scalable crowd-sourcing of video from mobile devices', 11th International Conference on Mobile systems, applications, and services, 2013, 139--152.





\bibitem{sun1}
Sun, Y., Guo, X., Zhou, S., Jiang, Z., Liu, X., \& Niu, Z., 'Learning-based task offloading for vehicular cloud computing systems', IEEE ICC, 2018, pp. 1--7.

\bibitem{sun}
Sun, Y., Zhou, S., \& Xu, J., 'Emm: Energy-aware mobility management for mobile edge computing in ultra dense networks', IEEE Journal on Selected Areas in Communications, 35(11), 2017, pp. 2637--2646.


\bibitem{wang}
Wang, T., Zhou, J., Liu, A., Bhuiyan, M. Z. A., Wang, G., \& Jia, W., 'Fog-Based Computing and Storage Offloading for Data Synchronization in IoT', IEEE Internet Things, 6, 2019, pp. 4272--4282.


\bibitem{wang1}
Wang, S., Urgaonkar, R., Zafer, M., He, T., Chan, K., \& Leung, K. K., ‘Dynamic service migration in mobile edge computing based on markov decision process’, IEEE/ACM Transactions on Networking, 27(3), 2019, pp. 1272—1288.


\bibitem{wang2}
Wang, S., Guo, Y., Zhang, N., Yang, P., Zhou, A., \& Shen, X., ‘Delay-aware microservice coordination in mobile edge computing: A reinforcement learning approach’, IEEE Transactions on Mobile Computing, 20(3), 2019, pp. 939—951.




\bibitem{xu1}
Xu, J., Ma, X., Zhou, A., Duan, Q., \& Wang, S., ‘Path selection for seamless service migration in vehicular edge computing’, IEEE Internet of Things, 7(9), 2020, pp. 9040--9049.

\bibitem{yang}
Yang, J., Zhang, H., Ling, Y., Pan, C., \& Sun, W., 'Task Allocation for Wireless Sensor Network Using Modified Binary Particle Swarm Optimization', IEEE Sensors Journal, 14(3), 2014, 882--892.

\bibitem{ref17}
Yao, Y., Cao, Q., \& Vasilakos, A. V., 'EDAL: An Energy-Efficient,Delay-Aware, and Lifetime-Balancing Data Collection Protocol for Wireless Sensor Networks', IEEE International Conference on Mobile Ad-Hoc and Smart Systems (MASS), 2013, 182--190.


\bibitem{zhang}
Zhang, Y., Zhao, J., Zheng, D., Deng, K., Ren, F., Zheng, X., \& Shu, J., ‘Privacy-preserving data aggregation against false data injection attacks in fog computing’, Sensors, 2018; 18(8): 2659.


\bibitem{ref21}
Zhou, A., Wang, S., Li, J., Sun, Q., \& Yang, F., 'Optimal  Mobile Device Selection for Mobile Cloud Service Providing', J. Supercomputing, 72(8):3222--3235, 2016.



\end{thebibliography}
\end{document}